\documentclass[a4paper,11pt]{article}
\pdfoutput=1 % if your are submitting a pdflatex (i.e. if you have
\usepackage{jinstpub} % for details on the use of the package, please
\usepackage[cp1250]{inputenc}
\usepackage{subfigure}
\usepackage{lineno}
\usepackage{amssymb}
\usepackage{float}
\usepackage{color}
\usepackage{graphicx}
\usepackage{rotating}
\usepackage{hyperref}
\usepackage{xcolor}
\usepackage{xspace}

\newcommand{\ardm}{ArDM}
\newcommand{\dart}{DArT}

\newcommand{\rn}{\ensuremath{\,\mathrm{^{222}Rn}}}

\newcommand{\ar}{\ensuremath{\,\mathrm{^{39}Ar}}}
\newcommand{\arb}{\ensuremath{\,\mathrm{^{37}Ar}}}
\newcommand{\arc}{\ensuremath{\,\mathrm{^{42}Ar}}}

\newcommand{\dia}{{DArT in ArDM}}

\newcommand{\keV}{\ensuremath{\,\text{ke\hspace{-.08em}V}}\xspace}

\title{DArT, a detector for measuring the \ar{}  depletion factor}

\author[a]{E.~Sanchez Garcia,}

\affiliation[a]{Centro de Investigaciones Energ\'{e}ticas, Medioambientales y Tecnol\'{o}gicas (CIEMAT), Av. Complutense 40, 28040 Madrid, Spain}

\emailAdd{edgar.sanchez@ciemat.es}

\abstract{One of the most powerful techniques for direct detection of dark matter via elastic scattering of galactic WIMPs is the use of   liquid argon time projection chambers. Atmospheric argon (AAr) has a naturally occurring radioactive isotope, \ar{}, of cosmogenic origin. The use of argon extracted from underground wells, deprived of \ar{}, is key to the physics potential of these experiments. The DarkSide-20k (DS-20k) dark matter search experiment will operate with 50 tonnes of radio-pure underground argon (UAr), extracted from the Urania plant in Cortez (USA) and purified in the Aria distillation plant (Sardinia, Italy). Assessing the radio-purity of UAr in terms of \ar{} is crucial for the success of DS-20k, as well as for future experiments of the Global Argon Dark Matter Collaboration (GADMC), and will be done with the experiment named \dia{}. DArT is a small chamber that will contain the  argon under test. The detector will be immersed in the LAr active volume of the ArDM detector, located at the Canfranc Underground Laboratory (LSC) in Spain, which will act as active veto for background events coming from photons   from  detector materials and  surrounding rock radioactivity. }


\collaboration[c]{on behalf of DarkSide collaboration}

\proceeding{LIDINE 2019: Light Detection in Noble Elements\\
28-30 August 2019\\
University of Manchester}

\begin{document}
\maketitle
\flushbottom

\section{Introduction}
\label{sec:intro}
Neutrino~\cite{EXO:2012,LBNO-DEMO:2014} and Dark Matter~\cite{DarkSide:2017,Aprile:2010,Darkside:2018} observation requires devices highly sensitive to tiny signals over large detecting volumes.  A large volume of a high purity noble element, like argon and xenon can provide, in principle, the sensitivity and detector-mass scalability required for those kinds of rare event searches. Additionally, the detection properties of liquid argon are particularly attractive for Dark Matter detectors, due to background rejection capability of signal-like nuclear recoils and electronic background using the different time distribution of the light output produced by these interactions. The DEAP-3600 experiment, with 3200~kg of LAr, has demonstrated an exceptional pulse shape discrimination (PSD) against such background, projected to be over $10^9$~\cite{Amaudruz:2017}.

Argon extracted from the atmosphere consists mostly in stable isotopes. However, three natural isotopes are produced by the interaction with the cosmic rays: \ar{},  \arb{} and  \arc{} ~\cite{Saldanha:2019}. The  \ar{} has the largest abundance and activity of the three ones. It produces a $\beta$ decay with a Q-value of 565~keV and its activity have been measured to be 1 Bq/kg~\cite{Loosli:1968}. For low background argon-based detectors  \ar{} is often the dominant source of interactions at low energies. Moreover, the pile-up and high rates produced by \ar{}  make difficult the operation of these detectors. 

The DarkSide-50 collaboration has measured the activity of underground argon (UAr) to be 0.73 mBq/kg~\cite{Darkside:2016}, equivalent to a 1400 depletion factor (DF) with respect to the AAr. For this reason, the next generation of Dark Matter direct detection experiments will operate with UAr in order to mitigate the effects of \ar{} . The DS-20k collaboration will extract at least 50 tonnes of UAr with the Urania plant. Urania will be able to deliver 330~kg/day of 99.99\% purity UAr. The UAr will be further chemically purified to detector-grade argon in Aria at the rate of 1~tonne/day. Aria is a 350~m cryogenic distillation plant currently being commissioned in Sardinia, Italy.

The \dia{}\ experiment  will measure the concentration of \ar{} in the UAr delivered by Urania and Aria and  will operate at the Canfranc Underground Laboratory (LSC, Spain). In the low background environment of the LSC, \dia{}  will be sensitive to very high depletion factors of \ar{}. 

\section{The \dia{} experiment}
\label{sec:experiment}
 DArT is a single-phase liquid argon detector, of about one litre active volume, that will be filled\ with underground argon samples. The light produced by the interactions in the active volume will be readout by two 1~cm$^2$ cryogenic silicon photomultipliers (SiPMs)~\cite{Acerbi:2017} that come, together with the readout electronics, from the DS-20k production chain.  An optical fiber reaching the inside of the chamber, illuminated with an LED, will be used  to monitor  the response of the SiPMs with time. 

\begin{figure}[!ht]
\centering
\includegraphics[width=0.85\linewidth]{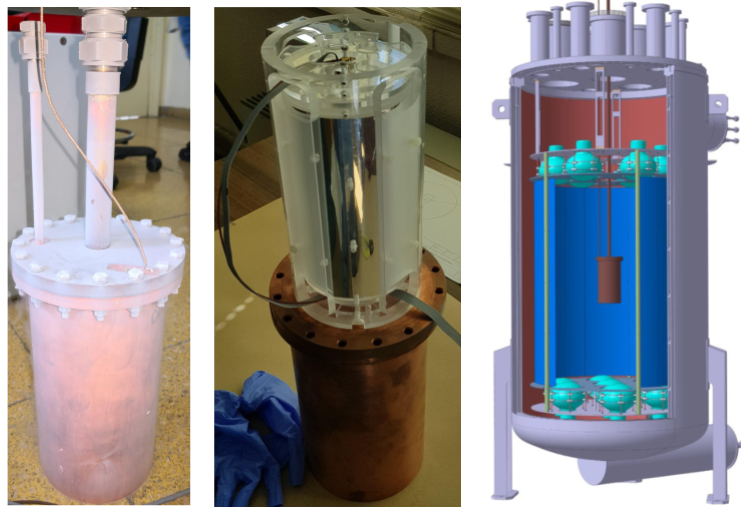}
\caption{ (Left) Photo of \dart\ vessel after leak test at cryogenic conditions.  (Center) Photo of acrylic inner support with the reflector around the active volume. The 2 PDM's are installed at top and bottom sides of the detector. (Right) Scheme of \dart\ inserted in the \ardm\ detector. }
\label{dart-experiment}
\end{figure}

\dart\ has been built with radio-pure materials to minimize the background induced by the intrinsic radioactive contamination of the detector. The outer vessel is made of ultra-pure 99.99\% Oxygen Free High Conductivity (OHFC) copper, with only 4 ppm of oxygen  (Fig.~\ref{dart-experiment}-Left). The inner structure is made of radio-pure acrylic (PMMA) and the surfaces  exposed to the argon are properly sanded to suppress  the plate-out of \rn\ daughters. The instrumentation and services of DArT are attached to the top flange, so that they are easily accessible by removing the copper vessel. Two 15~cm long copper pipes penetrate inside the vessel. A thinner pipe off axis, that goes all the way down the vessel, is used to fill the chamber with argon. A wider pipe, in the centre, stops at the top flange, allowing to route the necessary connections (SiPM HV, signal, level meter, etc.) through radio-pure cables.

The inner structure is made of 8 acrylic pieces (Fig.~\ref{dart-experiment}-Center): 1 attachment to the top cover, 1 outer and 1 inner cylinders, 2 SiPM supports, 2 end caps, and 1 facilities support on bottom. The two 6~mm thick disks are placed between the argon active volume and the two SiPMs. Both disks and the inner acrylic cylinder are coated with wavelength shifter (TPB) on its inner surface (200~$\mu$g/cm$^2$). The outer surface of the acrylic is enrolled in a Vikuiti reflector foil. An additional cylinder (annulus) is used as filler of the space outside the reflector foil. The outer acrylic parts is attached to the top vessel flange. Two dedicated boards, on top and bottom, attached to two acrylics disk using acrylic screws  hold the SiPMs used for reading out the light signals.

DArT will be placed in the centre of the 1~tonne liquid argon detector ArDM~\cite{ArDM:2017} (Fig.~\ref{dart-experiment}-Right), which will act as active veto against internal and external radiation. The single-phase setup of ArDM will be used, with a new set of low-radioactivity photo-multipliers (PMT), 6 on the top and 7 on the bottom, surrounded by a reflector foil covered with TPB. The whole setup is embedded in a 50~cm thick polyethylene shielding (not visible in the figure). In order to suppress the impact of  external gammas, which dominate the background budget, a 6~tonne lead shield will be installed around the ArDM vessel, in the hollow space between the ArDM cryostat and polyethylene shielding. The lead shield is a 140~cm height octagonal prism, with 10~cm width walls. In this configuration, the detector will be capable of measuring with high precision the larger \ar{} depletion factors of the UAr batches coming from the Urania plant.

\section{Detector performance}
\label{sec:sim}

In the \dart\ detector, signal events are electron recoils from the $\beta$ decay of \ar{}, which have an energy end-point of 565~\keV{}. These events deposit all the energy in \dart\, leaving no signal in the veto detector, \ardm.\ The range of the energy spectrum in DArT below 600~\keV{} is called \textit{region of interest} (ROI), as virtually all signal events pertain to it. Background events stem from radioactive decays in the detector materials and in the experimental hall surrounding the detector. They typically produce $\gamma$ particles that deposit energy in \dart\ and/or in \ardm.\ We assume that nuclear recoils are rejected using the pulse shape discrimination  technique, than in liquid argon is very powerful. The background events that leave a signal-like imprint in the \dart\ ROI, and an energy they deposit in \ardm\ above 10~\keV{}, are tagged as background and removed from the analysis. However, if the energy they deposit in \ardm\ is below that threshold, they cannot be tagged. These \textit{untagged} events are the background of the experiment.

\begin{figure}[!ht]
\centering
\includegraphics[width=0.49\linewidth]{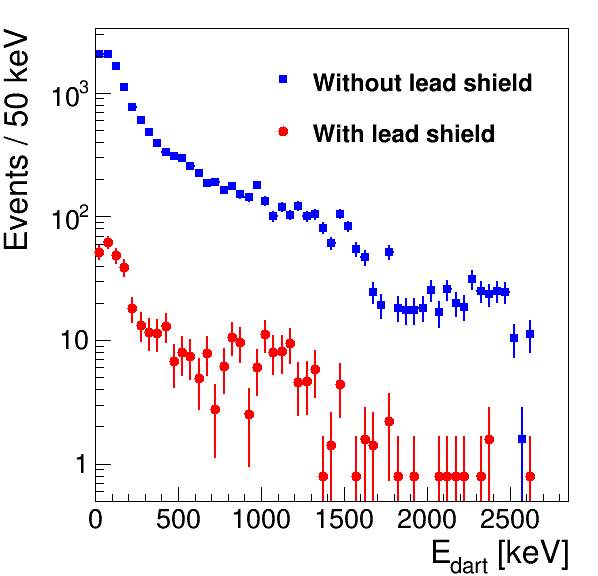}
\includegraphics[width=0.49\linewidth]{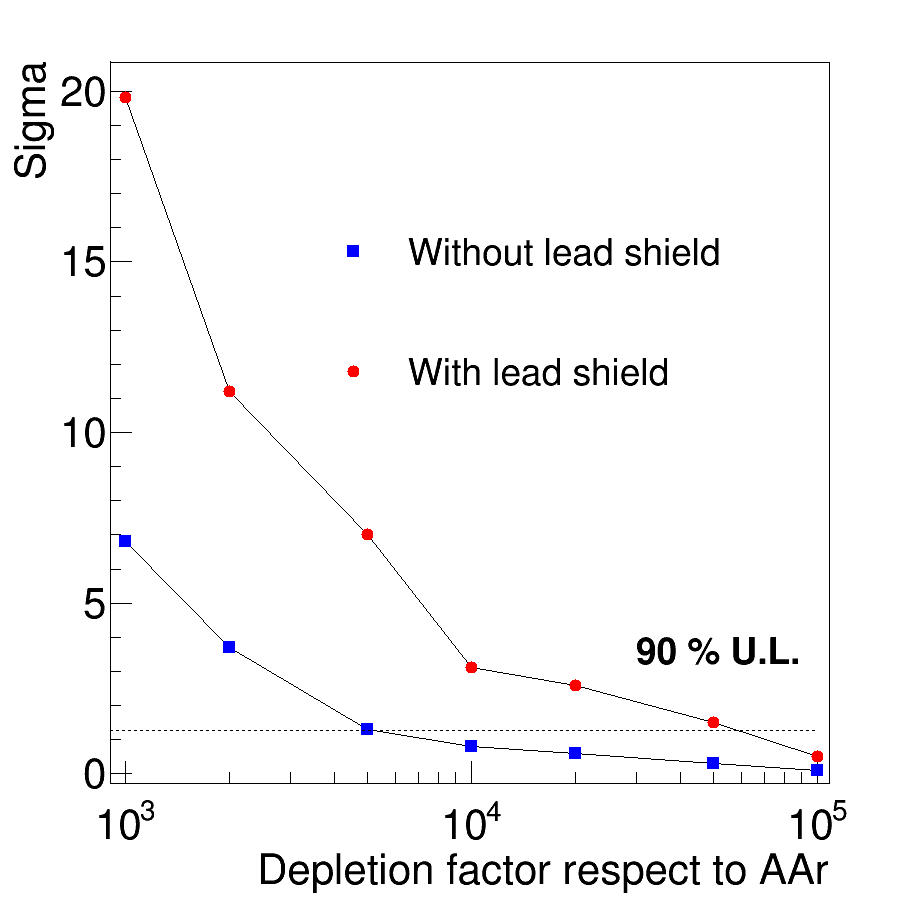}

\caption{(Left) Expected untagged background events in \dart{} with the lead shield (red) and without  (blue) in one month of data taking, assuming a veto threshold of 10~keV. (Right) Uncertainty expected per week for different argon depletion factors. The 90\% upper limit is indicated with a horizontal line. The red points correspond to the configuration with the lead shield and the blue ones without it.}
\label{dart-background}
\end{figure}

Extensive studies of the physics reach of the experiment are performed with tuned-on-data Monte Carlo simulations based on Geant4, with the simulation framework of the DarkSide-50 experiment (G4DS~\cite{Agnes:2017}). This package contains a detailed geometrical description of \dart\ and \ardm\, including the new lead shield. Predictions for the number of signal and background events expected in the ROI  are based on these simulations. For the background activity, actual data of the screened materials and measurements in the underground hall A of the LSC are used. The total number of background events per week in the ROI with the lead shield is 7210 out of which  465 are untagged,  around 35\% of these correspond to external background. For comparison, the expected number of background events in the ROI without a lead shield is 122000, with 10300 untagged events per week,  96\% is external background (Fig.~\ref{dart-background}-Left).

The energy deposits in the liquid argon of \dart{} produce VUV photons that, once converted to visible photons (420~nm) by the TPB, are either detected by the two SiPMs or lost in a dead area. The simulation of the propagation of photons to the SiPMs with Geant4 takes into account the optical properties of the detector materials and their interfaces, in particular the acrylic (PMMA), the TPB coating of the internal surfaces, the reflector foil and the SiPM planes.  Absorbed photons in the SiPMs planes are converted to photo-electrons, with a photon detection efficiency (PDE) of 40\%. Simulation parameters are tuned according to reference~\cite{Agnes:2017}).

The light response of the detector is evaluated simulating 20000 \ar{} events in the \dart\ active volume. The number of VUV-photons produced is 48 ph/\keV{}, which are converted into visible photons with an efficiency of 98.5\%. The reflector foil is placed on the sides of the active volume and  on the top and bottom acrylic caps, around the SiPMs. The average light collection efficiency is about 58\%, that corresponds to a light yield of 11 PE/\keV{}. Events closer to one of the detection planes have larger collection efficiency. The dependence of the efficiency on the radius of the interaction position is more pronounced for events close to the SiPMs, for geometrical reasons. However, the overall light collection efficiency is uniform.  The energy resolution is calculated with  monoenegetic deposits from 50~keV up to 800~keV uniformly distributed in the \dart\ active volume. An energy resolution between 3 \%  and  6 \%  is found in the ROI.

After the characterization of the background and the light response in \dart, we evaluate the sensitivity expected for different \ar{} depletion factors. The photoelectron spectrum for signal and background expected is generated using a lookup-table base in the  \ar{} events. The \ar{} content is extracted from a fit of the observed shape to the randomized weighted sum of the signal and background distributions. From these fits, the expected statistical uncertainties per week on the measurement of different depletion factors of \ar{} are shown in Fig.~\ref{dart-background}-Right. The 90\% one-sided confidence level upper limit (1.28 sigma)  is reached with a 5000 DF without the lead shield, this value is increased a factor 10 (50000 DF) with the lead shield.  Additionally, in the configuration with the lead shield, the uncertainty is below 1\% for a depletion factor of 10, 1\% for $DF = 100$, 7\% for $DF = 1400$ and 40\% for $DF = 14000$. Without the lead shied, the statistical uncertainties increase, typically, by a factor 3.

These results obtained assume the knowledge of the signal and background shape from the simulations and, therefore, do not include systematic uncertainties. Systematic uncertainties on the \ar{} decay spectrum can be significantly reduced validating the simulated spectra with a large data sample in an AAr run.

\section{Conclusions}
\label{sec:conc}

UAr is the key for the next generation Dark Matter experiments.  Also, neutrino experiments will benefit from high purity UAr. The Global Argon Dark Matter Collaboration has put 
in place a very ambitious program for the UAr production at large scale. The DArT in ArDM 
detector will play a fundamental role in this project. We have designed en experiment capable 
of measuring with high precision the residual amount of $^{39}$Ar in the argon of small samples. The 
performance of DArT has been studied with detailed background and optical simulations. These 
studies show that our detector would allow us to measure UAr depletion factors above 1000 with 
less than 10\% statistical precision. The construction of DArT is in a well-advanced stage and is 
expected to start operations at the beginning of 2020.

\acknowledgments
This research is funded by the Spanish Ministry of Economy and Competitiveness (MINECO) through the grant FPA2015-70657P. The author is also supported by the ``Unidad de Excelencia Mar\'{i}a de Maeztu: CIEMAT - F\'{i}sica de Part\'{i}culas'' through the grant MDM-2015-0509. 


\end{document}